\documentclass[11pt]{article}
\usepackage[margin=1in]{geometry}
\usepackage{amsmath,amssymb,amsthm}
\usepackage{hyperref}
\usepackage{enumitem}
\usepackage{authblk}
\usepackage{booktabs}
\usepackage{natbib}
\usepackage{array}
\usepackage[T1]{fontenc}
\hypersetup{
    colorlinks=true,
    linkcolor=blue,
    citecolor=blue,
    urlcolor=blue
}

\title{\textbf{Neural Manifolds as Crystallized Embeddings:\\
A Synthesis of the Free Energy Principle, Generalized Synchronization,\\
and Hebbian Plasticity}}

\author[1,*]{Vikas N. O'Reilly-Shah}
\affil[1]{Department of Anesthesiology \& Pain Medicine, University of Washington School of Medicine, Seattle, WA, USA}
\affil[*]{Correspondence: \texttt{voreill@uw.edu}}

\date{\today}

\begin{document}
\maketitle

\begin{abstract}
\noindent
The free energy principle casts perception as variational inference, but its biological implementation is underspecified. The generalized-coordinate formalism is not a literal claim that neurons compute arbitrary Taylor expansions. This paper argues that generalized synchronization (GS) provides the missing bottom-up mechanism. Certain recurrent circuits satisfy a contraction property: nearby trajectories converge exponentially toward each other. A contracting circuit driven by structured sensory input synchronizes to the driving dynamics. Under conditions that hold for all but a measure-zero set of circuit configurations, called generic embedding conditions, the resulting synchronization map embeds the low-dimensional sensory manifold into neural state space. The geometry predicted by the free energy principle is not imposed from above by an explicitly Bayesian neural calculus. It arises from ordinary recurrent dynamics driven by the world.

I then propose a developmental extension. Hebbian plasticity acting on the correlations generated by sensory-driven synchronization shapes the embedded manifold into recurrent connectivity, producing a continuous attractor network whose attractor approximates the embedded sensory manifold. Prediction-separation results from the companion paper \cite{ORS_Selvitella2026} bound the representational fidelity of the resulting circuit by prediction accuracy: where the network predicts future observations well, the synchronization map separates underlying states; where prediction fails, the representation collapses. The collapses are observable as categorical perception, metameric equivalence, and discrimination thresholds. On this view, mature head-direction, grid-cell, and stimulus-driven visual manifolds are developmental products of three interacting processes: dynamical contraction, generalized synchronization, and correlation-based plasticity. The central open problems are whether the Hebbian fixed point exists and whether Hebbian dynamics produce a sufficiently accurate predictor on the relevant input distribution.
\end{abstract}

\section{Introduction: The Bridge to Biological Mechanism}

The free energy principle (FEP), developed by Friston and colleagues across two decades, casts cortical computation as variational Bayesian inference in generalized coordinates of motion. This is a Taylor-series-style decomposition of sensory trajectories on which approximate posterior beliefs evolve via gradient descent on variational free energy \cite{Friston2010,Friston2010Generalised}. The framework is formally specified and has generated computational models of perception, action, and learning across multiple domains. Its biological implementation is not a literal correspondence: the generalized-coordinate formalism is not a literal claim that neurons explicitly compute arbitrary Taylor expansions, and gradient flows in generalized-coordinate space do not have direct correlates in identifiable neural mechanism.

Friston and colleagues have addressed this gap primarily through process theory. Predictive coding maps variational inference onto hierarchical message passing, with descending predictions and ascending prediction errors. Precision weighting is linked to synaptic gain and attention. Active inference connects perception to action through the fulfillment of sensory and proprioceptive predictions. Learning is treated as parameter optimization in a generative model. A Markov blanket is a statistical boundary that separates a system's internal states from its external environment; Markov-blanket formulations place perception, action, and self-organization within a broader account of biological systems at non-equilibrium steady state \cite{Friston2010,Friston2010Generalised,Palacios2019,Friston2021,Friston2023}. Within this biological bridge, generalized synchronization (GS) appears as the dynamical relation between internal and external states under free-energy minimization.

A parallel thread connects this synchronization picture to embedding theory. Kiebel, Daunizeau, and Friston \cite{KiebelDaunizeauFriston2009} compared generalized coordinates with Takens-style temporal embedding. Friston, Breakspear, and Deco \cite{FristonBreakspearDeco2012} formalized free-energy minimization as inducing generalized synchronization between brain and environment. Palacios et al.\ \cite{Palacios2019} extended this picture to coupled networks of mutually inferring neurons, and Friston et al.\ \cite{Friston2021} developed the Markov-blanket version in stochastic dynamical systems. Most recently, Friston et al.\ \cite{Friston2025} invoked Takens' theorem directly in a renormalizing active-inference setting, with an explicit appeal to Haken's synergetics and the slaving principle. These papers make an embedding-theoretic reading of FEP natural, but they do not supply the quantitative embedding theory or the developmental mechanism.

This paper makes that step, and the motivation for making it. I argue that generalized synchronization provides a bottom-up mechanism for the biological implementation of FEP. A contracting recurrent circuit driven by structured sensory input synchronizes to the driving dynamics. Under generic embedding conditions, the resulting synchronization map embeds the sensory manifold into neural state space. On this view, the geometry described by FEP need not be imposed by an explicitly Bayesian neural calculus. It arises from recurrent dynamics driven by the world, and is then stabilized developmentally by Hebbian plasticity acting on the activity correlations generated by that synchronization.

The developmental endpoint is the continuous attractor network whose attractor manifold approximates the embedded sensory manifold, sculpted developmentally by the conjunction of dynamical contraction and correlation-based plasticity rather than genetically prespecified. The empirical regularity that motivates the claim is a finding of twenty-first-century systems neuroscience: when the brain represents a sensory variable that lives on a low-dimensional geometric object, the corresponding neural population activity traces out that same geometric object. Head direction cells in the rodent thalamic and cortical head-direction circuit form a ring \cite{Chaudhuri2019,Peyrache2015}. Grid cells in entorhinal cortex form a torus \cite{Gardner2022,diSarra2025}. Population activity in mouse visual cortex traces a circular manifold under densely sampled stimulus rotations \cite{Beshkov2024}; Stringer et al.\ \cite{Stringer2019} provide complementary evidence on the high-dimensional geometry of V1 population responses. Across these and other systems, neural population activity recovers low-dimensional geometric structure corresponding to the represented variable, even when thousands of neurons participate.

Despite the regularity of the phenomenon, no unified mechanistic account exists for why cortical representation takes this geometric form. The continuous attractor network (CAN) literature \cite{KhonaFiete2022,BurakFiete2009} describes these manifolds as autonomous attractors of structured recurrent connectivity, but typically takes that connectivity as given by hand-designed templates or treats its developmental origin as a separate problem.

The free energy principle \cite{Friston2010} and predictive processing frameworks explain that cortex performs inference but, on their own, do not specify why the inference produces this particular geometry rather than some other. The neural manifold literature characterizes the geometry empirically but does not derive it from principle. In this paper, I argue that the pieces of a unified account exist and have not been assembled. Motivated by the bottom-up advantages of an embedding-theoretic view \cite{OReillyShah2025,OReillyShahNDPLS} and my other work in this space \cite{ORS_Selvitella2026}, the argument is developed across six sections. Section \ref{sec:fep} traces the points at which Friston and colleagues connect generalized coordinates, generalized synchronization, Takens-style embedding, Markov blankets, and self-organization. Section \ref{sec:reservoir} reviews the parallel development in reservoir computing and dynamical systems theory, where it has been formalized that generalized synchronization \emph{is} a topological embedding of the source attractor when ambient dimension exceeds twice the source-attractor dimension, a sufficient condition known as the generic Whitney embedding bound \cite{Whitney1936}; faithful distance-preserving (isometric) embedding is achievable under Nash's higher-dimensional bound. Section \ref{sec:plasticity} reviews the contraction-theoretic study of Hebbian-plastic recurrent networks. Section \ref{sec:synthesis} proposes the synthesis: Hebbian plasticity acting on the activity correlations produced by sensory-driven generalized synchronization shapes the embedded synchronization manifold into autonomous CAN connectivity, with representational fidelity bounded by prediction accuracy. Section \ref{sec:predictions} derives empirical predictions and identifies the open mathematical problems. Section \ref{sec:relations} discusses relations to neighboring frameworks: Haken's synergetics, slow feature analysis, self-organizing maps, and population coding theory.

Each ingredient appears in prior work. The contribution is to assemble the chain explicitly and identify which links are established and which remain open.

\section{The Free Energy Principle's Embedding Trajectory (2009--2025)}
\label{sec:fep}

The first move toward an embedding-theoretic account of cortical representation under the free energy principle was made in Kiebel, Daunizeau, and Friston \cite{KiebelDaunizeauFriston2009}. In a discussion of how variational filtering tracks fast sensory dynamics through generalized coordinates of motion, the authors wrote:
\begin{quote}
The use of generalised coordinates is formally similar to temporal embedding in the characterisation of dynamical systems: Takens' theorem \cite{Takens1981} states that it is possible to embed (i.e., geometrically represent) the structure of a vector-field in a higher dimensional space. This means that one can reconstruct the structure of the manifold, on which dynamics unfold, by using a Taylor expansion of the vector-field. This is very close to the idea of projecting the system into generalized coordinates.
\end{quote}
In the same paper, the authors speculated on synchronization: ``Coupled nonlinear systems naturally evolve towards a synchronous state, even with relatively weak coupling. It would be very interesting if these synchronised states could be associated with optimised free-energy states that are mandated by perception in particular and the free-energy principle in general.''

Friston, Breakspear, and Deco \cite{FristonBreakspearDeco2012} developed this connection formally. In a skew-product coupling, one subsystem evolves autonomously and drives the other without feedback. They treated perceptual dynamics as such a skew-product system, with environmental hidden states driving neuronal response states, and framed the resulting coupling in terms of generalized synchronization. They distinguished weak from strong synchronization based on the smoothness of the synchronization map: weak synchronization corresponds to a continuous $C^0$ but non-smooth map with fractal synchronization manifold and global dimension exceeding the driver dimension, while strong synchronization corresponds to a smooth $C^1$ map preserving dimensions. A Lyapunov exponent measures the average rate of exponential divergence or convergence of nearby trajectories; conditional Lyapunov exponents are computed while conditioning on a specific driving trajectory. They used conditional Lyapunov exponents to characterize the stability of the synchronization manifold, partitioning them into components tangential to the manifold (measuring stability along it) and transverse to it (measuring convergence toward it). They demonstrated through simulations of avian song perception that free-energy minimization produces fluctuations of conditional Lyapunov exponents around small near-zero values, a regime they termed self-organized critical slowing, in which the system sits near the boundary between stable and unstable behavior. The formal foundations cited were \cite{HuntOttYorke1997,Barreto2003}.

Palacios, Isomura, Parr, and Friston \cite{Palacios2019} extended the framework to coupled networks of mutually inferring neurons, demonstrating that emergent generalized synchronization arises directly from local free-energy minimization. The 2019 work also showed that synaptic plasticity in this setting produces structure learning that reflects the causal structure of exogenous input, although the plasticity rule used was free-energy gradient descent rather than Hebbian correlation. In a parallel paper, Palacios, Razi, Parr, Kirchhoff, and Friston \cite{Palacios2019b} developed the Markov-blanket and hierarchical self-organisation framework.

Friston, Heins, Ueltzh\"offer, Da Costa, and Parr \cite{Friston2021} took the framework into Markov-blanket territory. Using coupled stochastic Lorenz systems, they showed that conditional independencies at non-equilibrium steady-state induce a particular partition of states into internal, blanket, and external sets, and that the coupling between internal and external states across the blanket is characterized as a generalized synchrony or synchronization of chaos. They cited \cite{HuntOttYorke1997,Barreto2003} for the formal grounding of the synchronization claim.

Friston, Da Costa, Sakthivadivel, Heins, Pavliotis, Ramstead, and Parr \cite{Friston2023} provided the path-integral formulation. Encoding the dynamics of any random dynamical system by a Lagrangian playing the role of self-information, they showed that the most-likely internal path under free-energy minimization encodes a Bayesian belief about the most-likely external path. The synchronization between internal and external paths is interpreted as inference: the internal path parameterizes a conditional density over external paths.

Most recently, Friston, Heins, Verbelen, Da Costa, Salvatori, Markovic, Tschantz, Koudahl, Buckley, and Parr \cite{Friston2025} returned to embedding directly. In a section addressing how renormalizing generative models recover stochastic chaos from sequences of static images, the authors wrote: ``The answer lies in Takens' embedding theorem \cite{Takens1981,DeyleSugihara2011}, which means that any (chaotic) attractor can be reconstructed from a time-delay embedding, which is implicit in the temporal RG operators used in renormalization.'' They cited \cite{OrstavikStark1998} on coupled-map-lattice spatiotemporal embedding. The continuous-time link to the renormalization-group framework, they noted, is generalized coordinates of motion \cite{Friston2010Generalised}, with the renormalization-group formulation as the discrete homologue.

These papers articulate an account in which (i) free-energy minimization induces generalized synchronization between brain and environment; (ii) generalized coordinates of motion and renormalization-group operators implement a Takens-style temporal embedding; (iii) the synchronization manifold has the formal properties of a Bayesian belief about the world. They make the bridge to embedding theory natural, but they do not supply the quantitative dimensional content from the reservoir-computing line or the developmental mechanism proposed here.

What this trajectory has not done is quantify the embedding-theoretic content of the synchronization claim, nor use it for explanatory power. The $C^0/C^1$ distinction is invoked qualitatively. The dimension counting uses the Kaplan-Yorke formula rather than Whitney's bound. The Kaplan-Yorke formula estimates attractor dimension from the cumulative sum of the ordered Lyapunov exponent spectrum, applied here to conditional Lyapunov exponents. The Nash isometric embedding is not invoked. Although Friston et al.\ \cite{Friston2025} cite \O rstavik and Stark \cite{OrstavikStark1998} on coupled-map-lattice embedding, the specific forced and stochastic delay-embedding theorems of Stark \cite{Stark1999} and Stark et al.\ \cite{Stark2003}, the formal results that handle exactly the Langevin-driven skew-product structure on which the FEP rests, have not been developed as part of the FEP's perception and generalized-synchronization account. These are precisely the ingredients that the parallel reservoir-computing literature has developed.

\section{The Reservoir-Computing Formalization of GS as Embedding}
\label{sec:reservoir}

Independently of the FEP literature, the reservoir computing community has developed a quantitative formalization of generalized synchronization as an embedding. The relevant theorems are due to Hart and collaborators, building on Stark's prior work on delay embeddings for forced systems.

Stark \cite{Stark1999} and Stark, Broomhead, Davies, and Huke \cite{Stark2003} extended the classical Takens embedding theorem to forced systems. For a skew-product system in which a driver state $x$ evolves autonomously and a response state $y$ is forced by the driver, the time-delay embedding constructed from observations of the response generically (for all but a measure-zero set of system configurations) yields a topological embedding of the driver attractor into the reconstructed state space, provided the embedding dimension exceeds twice the dimension of the driver attractor and certain conditions on the periodic orbits of the forcing are satisfied. The 2003 paper extends the result to stochastic forcing, which is precisely the Langevin-driven structure that the FEP invokes.

A network satisfies the echo state property when its response is determined by the current and past inputs alone, independent of initial conditions. A reservoir network satisfying this property is called an echo state network (ESN). Hart, Hook, and Dawes \cite{HartHookDawes2020} showed that for such networks driven by structured input, the network admits an embedding of the input system's attractor under conditions on the contraction and observation function. Grigoryeva, Hart, and Ortega \cite{Grigoryeva2023} proved a more general result:
\begin{quote}
The celebrated embedding theorem of Takens is a particular case of a much more general statement according to which, randomly generated linear state-space representations of generic observations of an invertible dynamical system carry in their wake an embedding of the phase space dynamics into the chosen Euclidean state space. This embedding coincides with a natural generalized synchronization that arises in this setup and that yields a topological conjugacy between the state-space dynamics driven by the generic observations of the dynamical system and the dynamical system itself.
\end{quote}
A topological conjugacy is a continuous invertible map that identifies two dynamical systems as copies of each other, preserving all qualitative dynamics. Hart \cite{Hart2024} extended the result to continuous-time reservoir computers, providing a closed-form integral representation of the synchronization function in the linear case and establishing $C^1$ regularity under explicit contraction and observation conditions.

Hart \cite{Hart2025} proved the Whitney/Nash version. A second condition, called bunching, requires that the reservoir's contraction rate dominates the expansion rate of the driving dynamics along orbits; this domination ensures that the synchronization map is smooth, under the regularity and compactness hypotheses of Hart's theorem. For a generic state-update map $F$, among all smooth maps satisfying the echo state property and bunching, of reservoir dimension $N > 2q$ where $q$ is the dimension of the source attractor, the generalized synchronization function $f: \mathcal{M} \to \mathbb{R}^N$ is a $C^1$ embedding of the source attractor. The genericity is over the full space of smooth contracting maps, not restricted to linear reservoirs; this is the result that makes the account applicable to nonlinear biological circuits. Separately, for sufficiently high dimension satisfying Nash's bound, an isometric generalized synchronization exists; Hart \cite{Hart2025} constructs this explicitly when both the reservoir and source dynamics are linear.

Wong, Martin, and Eckhardt \cite{WongMartinEckhardt2024} identified the contraction conditions on continuous-time leaky reservoirs that guarantee the existence of generalized synchronization. A continuous-time leaky reservoir of the form $\dot{x} = -Cx + \sigma(Ax + Bu)$ is globally contracting when the connectivity matrix $A$ satisfies logarithmic-norm conditions relative to the leak matrix $C$. The logarithmic norm $\mu_p(A)$ of a matrix measures its maximum instantaneous expansion rate in the $\ell_p$ metric; $\mu_1$ encodes column diagonal dominance and $\mu_\infty$ encodes row diagonal dominance. Contraction requires these norms to be sufficiently negative to overcome the leak rate. These are biologically interpretable conditions on the structure of cortical microcircuits.

A driven recurrent network with appropriate contraction structure admits a unique synchronization function from environmental states to neural states. Under generic Whitney conditions, the synchronization function is a topological embedding of the environmental attractor into the neural state space. Under sufficient dimensionality, the synchronization function is an isometric embedding in the Nash sense. The contraction conditions required are biologically plausible.

The key papers canvassed here show little explicit bibliographic contact between the reservoir-computing and FEP literatures. The two traditions have developed in parallel, with substantial conceptual overlap but largely independent citation networks.

\section{Hebbian Plasticity and Contraction Structure in Cortical Microcircuits}
\label{sec:plasticity}

A third literature, partially separate from both the FEP and reservoir-computing lines, has studied how Hebbian and Hebbian-like correlation-based plasticity acts in recurrent networks with contraction structure. The relevant results combine biological plausibility with mathematical tractability.

Three results from two literatures form an almost-complete chain from plasticity to embedding. Kozachkov, Lundqvist, Slotine, and Miller \cite{Kozachkov2020} showed that anti-Hebbian plasticity drives the recurrent weight matrix toward symmetry: the anti-symmetric part of $W$ decays to zero. The anti-Hebbian rule applied to excitatory connections, with the learning rate sign that suppresses correlated activity, additionally drives $W$ toward negative semi-definiteness. The full Kozachkov result therefore produces global contraction in every direction, which is the opposite of the marginal-stability condition required for a continuous attractor along a tangent submanifold. This tension is taken up in Section \ref{sec:synthesis}; for the present chain, the relevant claim is that biologically realistic plasticity drives $W$ toward symmetry, and that further structural conditions select which symmetric matrices are reached.

Centorrino et al.\ \cite{Centorrino2023} proved that symmetric weight matrices admit sufficient conditions for Euclidean ($\ell_2$) contraction in Hopfield and firing-rate networks, with log-optimal rates. Euclidean $\ell_2$ contraction provides one natural route to the contraction and bunching hypotheses used in the GS embedding conditions of Hart \cite{Hart2024} and Wong, Martin, and Eckhardt \cite{WongMartinEckhardt2024}: the operator-norm contraction regime under which the synchronization function is $C^1$ and the embedding theorems apply. In a separate paper, Centorrino, Bullo, and Russo \cite{Centorrino2024} extended the analysis to the joint neural-synaptic system with dynamic Hebbian weights, proving non-Euclidean contractivity of the coupled $(h, W)$ dynamics in weighted $\ell_1$ norms.

The chain assembles as: (1) biological plasticity drives $W$ toward symmetry; (2) symmetric $W$ admits Euclidean contraction conditions; (3) Euclidean contraction plus regular driving yields smooth embedding. The remaining gap is a timescale separation argument. Kozachkov's symmetrization is dynamic: $W$ is changing. Centorrino et al.'s \cite{Centorrino2023} Euclidean result is for fixed symmetric $W$. The connection requires that after plasticity has approximately converged and $W$ is approximately symmetric and stationary, the Centorrino et al.\ \cite{Centorrino2023} conditions apply and the embedding theorems activate. This is biologically reasonable: it is the difference between learning a sensory representation (plasticity during development) and using it (approximately frozen weights during steady-state perception). Timescale separation justifies treating $W$ as approximately frozen after plasticity converges. The frozen $W$ must independently satisfy the Euclidean contraction conditions of \cite{Centorrino2023} for the embedding theorems to apply.

The bunching condition deserves explicit mention. Hart's embedding theorems require both contraction and bunching. Whether the Hebbian-symmetric $W$ that emerges from the plasticity-contraction chain satisfies bunching for biological input statistics is a separate question from $\ell_2$ contraction itself. The two conditions can be checked independently, and bunching is not implied by contraction.

A separate strand of work has shown that Hebbian-like plasticity on driven recurrent networks produces continuous attractor structure. Bernacchia \cite{Bernacchia2014} (preprint 2007 \cite{Bernacchia2007}) analyzed a binary recurrent network with Hebbian plasticity and showed that a continuous attractor emerges when the input distribution matches the tuning-curve distribution: experience matches sensory coding. Sch\"onsberg et al.\ \cite{Schonsberg2025} showed that a recurrent network model with continuous Hebbian plasticity produces diverse perceptual biases. The Hebbian-architecture-generation literature in reservoir computing \cite{CazaletsDambre2026} has shown that unsupervised Hebbian adaptation reshapes reservoir architecture from input correlations, improving downstream processing.

A fourth strand has shown that Hebbian-like learning rules produce structured recurrent circuitry. Vafidis, Owald, D'Albis, and Kempter \cite{Vafidis2022} showed that path-integration accuracy in head-direction ring attractor models emerges from Hebbian learning. More broadly, Eckmann, Young, and Gjorgjieva \cite{Eckmann2024} demonstrated that synapse-type-specific competitive Hebbian learning forms functional structured recurrent networks, illustrating the capacity of biological plasticity rules to generate organized connectivity.

Biologically realistic Hebbian plasticity operating on driven recurrent networks produces autonomous continuous attractor structure. What has not been established is the relationship between this attractor structure and the synchronization-function image of the driving sensory variable. The Hebbian-CAN literature treats the resulting attractors as templates, often hand-designed, that match the statistics of input but are not formally derived from an embedding theorem. The reservoir-computing-GS literature treats the synchronization function but does not address Hebbian sculpting.

\section{Synthesis}
\label{sec:synthesis}

The synthesis proposed here is the following. The mathematical chain has three links, drawing on published results from three literatures. First, biological plasticity mechanisms (specifically, anti-Hebbian excitatory plasticity, inhibitory Hebbian plasticity, and excitatory-inhibitory balance) drive the recurrent weight matrix toward symmetry and, once symmetric, produce contraction in the Euclidean ($\ell_2$) norm \cite{Kozachkov2020,Centorrino2023}. Second, Euclidean contraction in a driven recurrent network, together with regularity of the source dynamics and the bunching condition, guarantees the existence of a unique, smooth synchronization function $f: \mathcal{M} \to \mathbb{R}^N$ \cite{HartHookDawes2020,Hart2024}. Third, for generic observation functions and generic contracting networks with $N > 2 \dim(\mathcal{M})$, the synchronization function is a $C^1$ embedding \cite{Hart2025}. Each link is established in the literature under additional hypotheses; the full composition also requires showing that intermediate weights along the developmental trajectory satisfy the contraction and bunching conditions the embedding theorems require, that the convergence endpoint $W^*$ develops the marginal tangent-transverse eigenstructure required for CAN behavior, and that these conditions hold jointly. This composition has not been assembled.

Consider a recurrent cortical circuit at a stage of the developmental trajectory when the weights have become approximately symmetric under Hebbian plasticity and satisfy the Euclidean contraction conditions of \cite{Centorrino2023}, driven by sensory input that lives on a low-dimensional environmental manifold. By Hart \cite{Hart2024,Hart2025}, the network admits a unique generalized synchronization function $f: \mathcal{M} \to \mathbb{R}^N$ that is a $C^1$ embedding of the environmental manifold $\mathcal{M}$ into the neural state space $\mathbb{R}^N$, provided $N > 2 \dim(\mathcal{M})$, and an isometric embedding under sufficient additional dimensionality. On the timescale of fast neural dynamics, the network state tracks $f(x_t)$ where $x_t$ is the current environmental state.

Now superimpose Hebbian plasticity on the recurrent connectivity. The activity correlations produced by the synchronization regime, integrated over experience, drive the recurrent weight matrix toward a fixed point. I conjecture that the Hebbian fixed point is a recurrent connectivity matrix whose autonomous attractor manifold approximates the image $f(\mathcal{M})$ of the synchronization function: the embedded environmental manifold. At this fixed point, the circuit no longer requires sensory drive to maintain the manifold; the manifold has been encoded into connectivity. The trained circuit is a continuous attractor network of the canonical neural type, whose attractor geometry approximates the embedded sensory manifold, constructed developmentally from the conjunction of generalized synchronization and Hebbian plasticity. I call this developmental process crystallization: the encoding of synchronization-function geometry into autonomous recurrent connectivity by Hebbian plasticity acting on sensory-driven activity correlations.

\subsection{The Tangent--Transverse Problem}

The contraction that licenses generalized synchronization and the marginal stability that defines a continuous attractor are opposite dynamical conditions. During the driven phase, the network contracts: all trajectories converge toward the synchronization-function image $f(\mathcal{M})$. In the autonomous CAN phase, the network must be marginally stable along $f(\mathcal{M})$ (permitting persistent activity at any point on the manifold) while contracting transverse to $f(\mathcal{M})$ (so that perturbations off the manifold decay). The synthesis requires that Hebbian plasticity, acting on the correlations of the GS-driven activity, weakens contraction specifically in the directions tangent to $f(\mathcal{M})$ while preserving contraction in the transverse directions.

This is a strong structural requirement. In correlation-based Hebbian models, the Hebbian fixed point $W^*$ depends on the second-order statistics $C = \langle f(x) f(x)^\top \rangle$ of the synchronization-driven activity under the input distribution. For the autonomous dynamics generated by $W^*$ to be marginally stable along $f(\mathcal{M})$ and contracting transverse to it, the Jacobian of those dynamics must have near-zero eigenvalues in the tangent directions of $f(\mathcal{M})$ and negative eigenvalues in the transverse directions. In the identity-gain linear-activation limit, the Jacobian reduces to $-I + W^*$, and the condition becomes a requirement on $W^*$'s eigenstructure directly. For curved $f(\mathcal{M})$ (a ring nonlinearly embedded in $\mathbb{R}^N$, a torus, an orientation hypersphere), no fixed $d$-dimensional leading eigenspace of $-I + W^*$ can equal the $d$-dimensional tangent space at every point. In the identity-gain linear-activation limit, the alignment is therefore approximate; nonlinear activation makes the Jacobian state-dependent; whether this improves alignment is open.

The mismatch between $W^*$'s autonomous attractor and the exact image $f(\mathcal{M})$ is representational error. That error is the substance of what the crystallization account predicts about perception.

\subsection{Approximation as a Feature: Prediction Accuracy Constrains Representation}

The synthesis requires the resulting circuit to be a sufficiently accurate predictor of future observations, not an exact reproduction of $f(\mathcal{M})$. Where the predictor is accurate, the synchronization map must separate the underlying states that produced those observations. Where the predictor fails, the representation collapses, and those collapses are observable as familiar perceptual phenomena.

The companion preprint \cite{ORS_Selvitella2026} establishes this quantitatively. Proposition 4.5 of that paper (the prediction-separation link) states the following. Let $\Gamma_K: \mathcal{M} \to \mathbb{R}^{K+1}$ be the $K$-step forward observation map, $\Gamma_K(x) = (\omega(x), \omega(\phi(x)), \ldots, \omega(\phi^K(x)))$, where $\phi$ denotes the dynamics on $\mathcal{M}$ and $\omega: \mathcal{M} \to \mathbb{R}$ is a scalar observation function. Suppose $\Gamma_K$ has separation modulus $\eta$: distinct states $x, x'$ at base-manifold distance at least $\delta$ have $\|\Gamma_K(x) - \Gamma_K(x')\| \geq \eta(\delta)$. Suppose a predictor $P: \mathbb{R}^N \to \mathbb{R}^{K+1}$ achieves uniform error $\sup_x \|P(f(x)) - \Gamma_K(x)\| \leq \varepsilon$. Then the synchronization map $f$ cannot collapse any pair of states whose base-manifold separation exceeds the threshold determined by $2\varepsilon < \eta(\delta)$:
\[
f(x) = f(x') \implies d_\mathcal{M}(x, x') < \delta.
\]
With $P$ additionally Lipschitz, Corollary 4.6 of \cite{ORS_Selvitella2026} gives a quantitative lower bound on $\|f(x) - f(x')\|$ in terms of $\|\Gamma_K(x) - \Gamma_K(x')\|$, $\varepsilon$, and the Lipschitz constant.

A good predictor cannot afford to collapse distinguishable futures. If the network's readout from $f$ can reconstruct the next $K$ observations to within $\varepsilon$, then states whose future observations differ by more than $2\varepsilon$ must map to distinct points in neural state space. The prediction objective constrains the representation.

\paragraph{Uniform versus expected error.} The proposition is stated for uniform error. Real learning systems minimize expected error under the input distribution. Distinctions the organism encounters frequently are predicted well and represented finely; rare distinctions are predicted poorly and collapse. This is categorical perception: a continuum of physical stimuli perceived as falling into discrete categories, with better discrimination across category boundaries than within them \cite{Harnad1987}. The same logic predicts expertise effects. Musicians show enhanced pitch discrimination relative to non-musicians \cite{Micheyl2006,Bidelman2011}. Trained oenologists make fine-grained discriminations among wines that novices perceive as equivalent \cite{Smith2007-oe}, an expertise effect that Dolega, Mentec, and Cleeremans \cite{Dolega2025-bl} situate within the quality space account. Learning reshapes the geometry of the representational space, increasing separation along task-relevant dimensions.

\paragraph{Variation in $\eta$.} The separation modulus of $\Gamma_K$ varies across the driving manifold. States whose $K$-step futures are nearly identical produce small $\eta$, regardless of how far apart they are on the base manifold. States whose $K$-step futures are near-identical have no guaranteed separation in the representation; whether they actually collapse depends on the specific synchronization map. In color vision, the special case where the observation function $\omega$ itself collapses distinct states (different spectral distributions producing indistinguishable cone responses) is the classical metamer. The crystallization account predicts an analogous collapse whenever $K$-step futures are near-identical, encompassing but extending beyond this classical case.

\paragraph{Finite $\varepsilon$.} Finite prediction error sets a hard resolution limit. The proposition guarantees separation only for states whose future observations differ by more than $2\varepsilon$. Below that scale, the proposition provides no separation guarantee. This is a discrimination threshold: the minimum physical difference the system resolves.

\begin{table}[ht]
\centering
\renewcommand{\arraystretch}{1.2}
\begin{tabular}{@{}>{\raggedright}p{5.5cm} p{6.5cm}@{}}
\toprule
\textbf{Limitation of the guarantee} & \textbf{Perceptual consequence} \\
\midrule
Uniform versus expected prediction error & Categorical perception, expertise effects \\
Variation in separation modulus $\eta$ & Metameric collapse \\
Finite resolution set by $\varepsilon$ & Discrimination thresholds \\
\bottomrule
\end{tabular}
\caption{Limits of the prediction-separation guarantee and their perceptual consequences.}
\label{tab:gaps}
\end{table}

The crystallization conjecture, in its softened form, is therefore: the Hebbian fixed point $W^*$ produces a circuit whose autonomous attractor approximates $f(\mathcal{M})$ with representational error bounded by Proposition 4.5 of \cite{ORS_Selvitella2026}. The exact-coincidence version of the conjecture would conflict with the curved-manifold obstruction described above; the approximate version respects the geometric constraints of the problem and yields falsifiable predictions about which states the circuit will and will not distinguish.

A second open question is left by this framing. The proposition shows that a good predictor must preserve separation, but it does not show that Hebbian dynamics produce a good predictor. The Hebbian rule minimizes a correlation-based objective, not a prediction-error objective. Whether the two objectives have nearby (or coincident) fixed points is a separate question, with partial answers in the information-maximization (infomax) \cite{Linsker1988} and predictive-coding-by-Hebbian-rule \cite{RaoBallard1999} literatures. The synthesis depends on this second link as much as on the first.

\subsection{Position Relative to Prior Work}

Each ingredient appears in prior work: Bernacchia \cite{Bernacchia2014} showed Hebbian-induced continuous attractors but without a synchronization-function framing; Lu and Bassett \cite{LuBassett2020} showed that invertible generalized synchronization combined with output-feedback training produces autonomous attractors matching source dynamics, but with First-Order Reduced and Controlled Error (FORCE) training rather than Hebbian dynamics on internal weights; Sch\"onsberg et al.\ \cite{Schonsberg2025} showed plastic-attractor formation under continuous Hebbian dynamics in driven recurrent networks; the Hebbian-architecture-generation literature \cite{CazaletsDambre2026} showed Hebbian sculpting of reservoir connectivity from input correlations. Palacios, Isomura, Parr, and Friston \cite{Palacios2019} combined synaptic plasticity with FEP-induced generalized synchrony, but with free-energy gradient plasticity rather than Hebbian correlation, and without the embedding-theoretic framing.

The synthesis developed in Section~\ref{sec:synthesis} assembles these results: a CAN whose attractor manifold approximates the topologically (and under sufficient dimension, isometrically) embedded sensory manifold, with approximation error governed by the prediction-separation link of \cite{ORS_Selvitella2026}.

The synthesis has biological content distinct from its predecessors. It does not require genetic prespecification of ring or toroidal connectivity; the connectivity is the developmental product of sensory experience filtered through generic contraction structure and Hebbian plasticity. Empirical predictions follow in Section \ref{sec:predictions}.

A scope note. Proposition 4.5 of \cite{ORS_Selvitella2026} is stated for autonomous driver dynamics $\phi$. The synthesis therefore covers perception under structured sensory drive. Active inference, where action closes the sensorimotor loop and the driver becomes policy-dependent, requires an extension beyond the strict skew-product setting and is not addressed here.

\section{Empirical Predictions and the Open Mathematical Problems}
\label{sec:predictions}

Five empirical predictions follow directly from the synthesis.

\paragraph{Prediction 1 (Generic embedding threshold).} For a $d$-dimensional sensory manifold, the reservoir-embedding results predict a generic sufficient ambient dimension threshold of $N > 2d$ for faithful embedding by generic observation/reservoir maps. Since $N$ is integer-valued and $d$ is the integer dimension of the sensory manifold, this corresponds to $N \geq 2d+1$. This is not an upper bound on the intrinsic dimension of the represented manifold, which remains $d$, nor is it a topological lower bound: special embeddings occur in lower ambient dimensions, such as $S^1 \subset \mathbb{R}^2$ and $T^2 \subset \mathbb{R}^3$. It is the generic Whitney/Takens-style threshold at which faithful recovery becomes stable under generic projections and adequate sampling.

\paragraph{Prediction 2 (Topological recovery across projection dimension).} When persistent homology is applied to population activity projected into increasing ambient dimension, the predicted Betti numbers become more reliably recoverable once the projection dimension reaches the generic embedding threshold $N > 2d$, assuming adequate sampling, signal-to-noise, and metric choice. Below this threshold, recovery fails or becomes unstable, although special low-dimensional embeddings preserve topology. For head-direction systems ($d = 1$), $\beta_1 = 1$ stabilizes at $N \geq 3$; for grid systems ($d = 2$), $\beta_1 = 2$ and $\beta_2 = 1$ at $N \geq 5$; for V1 orientation ($d = 1$), $\beta_1 = 1$ at $N \geq 3$.

\paragraph{Prediction 3 (Discrimination tracks embedding resolution).} The synchronization-function image is approximate. States whose images diverge slowly in neural state space are perceptually confusable; states whose images diverge quickly are discriminable. The psychometric function on any stimulus dimension tracks the local separation modulus of the synchronization function image, as bounded by Proposition 4.5 of \cite{ORS_Selvitella2026}. Categorical perception, metameric equivalence, and the structure of psychophysical thresholds emerge as consequences of embedding quality rather than requiring separate explanations. Frequently encountered distinctions are predicted finely and represented finely; rarely encountered distinctions collapse.

\paragraph{Prediction 4 (Developmental sensitivity).} Disrupting correlation-based recurrent plasticity during the formation of a continuous attractor network prevents autonomous attractor formation despite preserved sensory drive. Early sensory experience determines attractor geometry in cases where sensory statistics are manipulable.

\paragraph{Prediction 5 (Geometry tracks input statistics).} If sensory statistics are altered while plasticity is preserved, the resulting CAN attractor manifold tracks the altered statistics. This is a sharper version of Bernacchia's matching condition \cite{Bernacchia2014}, with the additional content that the matching is mediated by the prediction-separation link: the attractor manifold reflects which distinctions the input distribution makes available to the predictor.

The published empirical record on Predictions 1 and 2 matches the synthesis. The Betti numbers found in Chaudhuri et al.\ \cite{Chaudhuri2019}, Gardner et al.\ \cite{Gardner2022}, and di Sarra et al.\ \cite{diSarra2025} match those the synthesis predicts, at dimensions exceeding the Whitney bound. Sansford, Whiteley, and Rubin-Delanchy \cite{SansfordWhiteleyRubinDelanchy2025} revisit the Gardner et al.\ toroidal topology finding and establish that ambient intrinsic dimension much greater than $\log n$ suffices for persistence diagrams to reveal latent homology, a result that supports the Whitney-threshold prediction without directly establishing it. A formal mathematical treatment of the smooth-embedding regime in driven recurrent networks, including Proposition 4.5, is in the companion preprint \cite{ORS_Selvitella2026}.

\paragraph{Open mathematical problem 1: existence of the Hebbian fixed point.} The first open question is whether the Hebbian plasticity dynamics on the synchronization-driven activity admit a fixed point $W^*$, and whether intermediate weights along the developmental trajectory satisfy the Euclidean ($\ell_2$) contraction conditions of \cite{Centorrino2023} at each approximately frozen stage, and whether the convergence endpoint $W^*$ develops the marginal tangent-transverse eigenstructure required for CAN behavior. The technical ingredients exist: plasticity-driven symmetrization \cite{Kozachkov2020}, Euclidean contraction under symmetric weights \cite{Centorrino2023}, non-Euclidean contraction of the joint neural-synaptic system \cite{Centorrino2024}, the existence of synchronization functions under contraction \cite{Hart2024,WongMartinEckhardt2024}, and generic embedding under Whitney bounds \cite{Hart2025,Grigoryeva2023}. The composition is not obvious, precisely because of the metric mismatch between Centorrino et al.'s \cite{Centorrino2024} non-Euclidean contraction of the dynamic system and Centorrino et al.'s \cite{Centorrino2023} Euclidean contraction of the frozen system, and because Kozachkov's negative-semi-definite endpoint conflicts with the marginal stability the CAN requires.

I conjecture that under: (i) timescale separation between fast neural dynamics and slow plasticity, so that the embedding analysis applies to approximately frozen weights; (ii) an anti-Hebbian or mixed Hebbian rule that drives the weight matrix toward symmetry \cite{Kozachkov2020} but stops short of global negative-semi-definiteness, preserving a marginal mode (no existing result establishes when or why symmetrization halts before full negative-semi-definiteness; this is an assumption of the conjecture); (iii) the symmetric weights along the developmental trajectory (before convergence to $W^*$) satisfying the Euclidean contraction conditions of \cite{Centorrino2023}, so that the embedding theorems apply at each approximately frozen intermediate weight matrix; (iv) the Hebbian plasticity dynamics on the synchronization-driven activity converge to a fixed point $W^*$ (convergence is an additional assumption, not a consequence of conditions (i)--(iii)); and (v) the time-averaged activity correlation matrix $C = \langle f(\phi^t(x))\, f(\phi^t(x))^\top \rangle_t$ has its leading eigenspace approximately aligned with the tangent directions of $f(\mathcal{M})$, with the quality of this alignment depending on the curvature of $f(\mathcal{M})$ and the input distribution: $W^*$ supports an approximation of $f(\mathcal{M})$ as an autonomous attractor of the undriven dynamics, with near-zero Jacobian eigenvalues along the leading eigenspace of $C$ and negative eigenvalues in the complementary directions. The metric distortion between the autonomous attractor and the embedded sensory manifold depends on the local condition number of $f$ and the alignment error between $C$'s leading eigenspace and the tangent bundle of $f(\mathcal{M})$. Conditions (i)–(v) are the hypothesized sufficient conditions; no existing result proves that they entail the conclusion. Showing that they do, or finding a counterexample, is the content of Open Problem 1.

\paragraph{Open mathematical problem 2: Hebbian dynamics as predictor.} The second open question is whether the Hebbian fixed point produces a sufficiently accurate predictor of future observations. Proposition 4.5 of \cite{ORS_Selvitella2026} shows that prediction accuracy bounds representational fidelity, but it does not establish that Hebbian dynamics produce a good predictor. The Hebbian rule minimizes a correlation-based local objective. Prediction-error minimization is a different objective. The infomax line \cite{Linsker1988,BellSejnowski1995} and the Hebbian-implementation-of-predictive-coding line \cite{RaoBallard1999} have argued that these objectives have related fixed points under specific input statistics, but the relationship is not general. Settling whether the correlation-based Hebbian fixed point coincides with, or is close to, a prediction-error-minimizing fixed point on biologically realistic input distributions is the second prong of the open problem.

The conjunction of Open Problems 1 and 2 is the central technical question of the synthesis.

\paragraph{Technical gaps in the supporting chain.} Several smaller gaps remain. First, the embedding theorems of Hart \cite{HartHookDawes2020,Hart2025} are proved in discrete time; the FEP's generalized filtering operates in continuous time. Hart \cite{Hart2024} and Wong et al.\ \cite{WongMartinEckhardt2024} provide continuous-time GS results, but the full chain from contraction through smoothness to generic embedding has not been assembled in continuous time; Hart \cite{Hart2025} flags this as open. Second, the FEP involves stochastic (Langevin) dynamics; Stark et al.\ \cite{Stark2003} prove delay embeddings under stochastic forcing, but the GS-to-embedding chain under noise is less developed. Third, the genericity in Hart \cite{Hart2025} (``generic $F$ among all smooth contracting maps'') does not directly imply that any specific parameterized family of networks (such as tanh recurrent neural networks (RNNs) or biological firing-rate models) lies in the generic set. The expressiveness of these families makes this plausible, but no general proof is available. Fourth, the Euclidean contraction conditions of \cite{Centorrino2023} are proved for autonomous networks; in the driven case the $h$-Jacobian is unaffected by scalar input (the driving term enters additively), but careful statement is needed for vector-valued sensory input. Fifth, whether Friston's generalized filtering dynamics satisfy the contraction/echo-state conditions that license the embedding chain has not been proved; whether the precision-weighting mechanism provides sufficient contraction is open. Sixth, the bunching condition required by Hart's embedding theorems has not been verified for biologically realistic input statistics on Hebbian-symmetric $W$.

\section{Relation to Other Frameworks}
\label{sec:relations}

The synthesis connects to several adjacent traditions.

\paragraph{Haken's synergetics.} The synchronization function is the dynamical-systems heir of Haken's slaving principle \cite{Haken1977,Haken1996}, with fast neural dynamics enslaved by slow sensory order parameters on a low-dimensional manifold. The Haken-Kelso-Bunz (HKB) model \cite{HakenKelsoBunz1985} is the classical coordination-dynamics instance of slaving on $S^1$. Friston's 2025 paper \cite{Friston2025} makes this lineage explicit, dedicating itself to Haken's memory and noting that recursive application of the slaving principle is the synergetic foundation of FEP renormalizability. The crystallization claim adds a developmental dimension to the slaving lineage: the slaved configuration is sculpted into autonomous recurrent connectivity by Hebbian plasticity, and the slaved manifold survives, in approximate form, even when the driver is removed.

\paragraph{Self-organizing maps and slow feature analysis.} Self-organizing maps \cite{Kohonen1982,vonderMalsburgWillshaw1976} share the intuition that competitive plasticity from driven activity yields a topology-preserving map from input space into neural space. Slow feature analysis \cite{WiskottSejnowski2002,Franzius2007} produces place-cell- and head-direction-cell-like representations from temporal slowness. The crystallization synthesis differs from both. The synchronization function arises from the recurrent dynamics of the driven circuit, not a hand-imposed competitive lattice. The post-Hebbian attractor manifold is autonomous: the trained circuit no longer needs sensory drive to maintain it, and the dimensional content is set by Whitney/Nash bounds rather than by lattice geometry.

\paragraph{Population coding theory.} Rate coding, spike-timing coding, and population coding emerge as projections of the synchronization function onto different measurement axes. In the crystallization account, the conceptual dimensions of representation identified by Walker et al.\ \cite{Walker2023} and Pohl et al.\ \cite{Pohl2026} (sensitivity, specificity, invariance, and functionality) map onto properties of the synchronization function: continuity, injectivity, low-dimensional domain, and the prediction-separation link.

\paragraph{Predictive processing.} Predictive processing specifies the inferential structure of cortical computation. The crystallization synthesis specifies how that inferential structure is built developmentally and what geometry it produces. The two accounts are compatible; the synthesis adds content the predictive-processing account does not supply. Active inference, which closes the sensorimotor loop, requires an extension of the embedding theorems beyond the strict skew-product setting. The forced and stochastic delay-embedding theorems of Stark \cite{Stark1999,Stark2003} handle exactly the Langevin-driven skew-product structure the FEP invokes, but these results have not been incorporated into the FEP's generalized-synchronization account (as noted in Section~\ref{sec:fep}).

\paragraph{Bayesian mechanics.} The recent Bayesian-mechanics formulations of the FEP \cite{Sakthivadivel2022,Ramstead2023} provide axiomatic foundations for the variational structure of self-organizing systems. The crystallization synthesis gives a biophysical interpretation of Bayesian mechanics in cortical circuits: the abstract variational structure is realized by contraction-induced synchronization, Hebbian-induced approximation of the synchronization manifold, and the prediction-separation link acting together.

\section{Conclusion}

This synthesis assembles three lines of work that developed in largely separate citation networks: the free energy principle's synchronization trajectory, the reservoir-computing formalization of generalized synchronization as topological embedding, and the contraction-theoretic analysis of Hebbian-plastic recurrent networks. The assembly produces a testable mechanistic account of neural manifold formation, with empirical predictions specified in Section~\ref{sec:predictions} and two open mathematical problems identified there. Until those problems are resolved, the crystallization account is a supported conjecture rather than a proved chain.

The account shifts the developmental question. Which sensory statistics and plasticity dynamics produce a given geometry becomes the question, rather than which prespecified connectivity template does. The representational errors of the resulting circuit are themselves the perceptual phenomena the crystallization account predicts.

\section*{Acknowledgments}

I thank Alessandro Selvitella for extensive prior collaboration on the formal mathematical aspects of generalized synchronization in driven recurrent networks, the substance of which appears in the companion preprint \cite{ORS_Selvitella2026}. The views expressed here, and any errors of synthesis, are mine alone.

\bibliographystyle{plain}
\bibliography{refs}

\end{document}